\documentstyle[12pt]{article}

\begin{document}

\def \ajp#1#2#3{Am. J. Phys. {\bf#1}, #2 (#3)}
\def \apny#1#2#3{Ann. Phys. (N.Y.) {\bf#1}, #2 (#3)}
\def \app#1#2#3{Acta Phys. Polonica {\bf#1}, #2 (#3)}
\def \arnps#1#2#3{Ann.~Rev.~Nucl.~Part.~Sci.~{\bf#1}, #2 (#3)}
\def \arns#1#2#3{Ann.~Rev.~Nucl.~Sci.~{\bf#1}, #2 (#3)}
\def \cn{Collaboration}
\def \cp89{{\it CP Violation,} edited by C. Jarlskog (World Scientific,
Singapore, 1989)}
\def \efi{Enrico Fermi Institute Report No. EFI}
\def \hb87{{\it Proceeding of the 1987 International Symposium on Lepton and
Photon Interactions at High Energies,} Hamburg, 1987, ed. by W. Bartel
and R. R\"uckl (Nucl. Phys. B, Proc. Suppl., vol. 3) (North-Holland,
Amsterdam, 1988)}
\def \ib{{\it ibid.}~}
\def \ibj#1#2#3{~{\bf#1}, #2 (#3)}
\def \ichep72{{\it Proceedings of the XVI International Conference on High
Energy Physics}, Chicago and Batavia, Illinois, Sept. 6 -- 13, 1972,
edited by J. D. Jackson, A. Roberts, and R. Donaldson (Fermilab, Batavia,
IL, 1972)}
\def \ijmpa#1#2#3{Int. J. Mod. Phys. A {\bf#1}, #2 (#3)}
\def \ite{{\it et al.}}
\def \lkl87{{\it Selected Topics in Electroweak Interactions} (Proceedings of
the Second Lake Louise Institute on New Frontiers in Particle Physics, 15 --
21 February, 1987), edited by J. M. Cameron \ite~(World Scientific, Singapore,
1987)}
\def \ky85{{\it Proceedings of the International Symposium on Lepton and
Photon Interactions at High Energy,} Kyoto, Aug.~19-24, 1985, edited by M.
Konuma and K. Takahashi (Kyoto Univ., Kyoto, 1985)}
\def \mpla#1#2#3{Mod. Phys. Lett. A {\bf#1}, #2 (#3)}
\def \nc#1#2#3{Nuovo Cim. {\bf#1}, #2 (#3)}
\def \np#1#2#3{Nucl. Phys. {\bf#1}, #2 (#3)}
\def \pisma#1#2#3#4{Pis'ma Zh. Eksp. Teor. Fiz. {\bf#1}, #2 (#3) [JETP Lett.
{\bf#1}, #4 (#3)]}
\def \pl#1#2#3{Phys. Lett. {\bf#1}, #2 (#3)}
\def \plb#1#2#3{Phys. Lett. B {\bf#1}, #2 (#3)}
\def \pr#1#2#3{Phys. Rev. {\bf#1}, #2 (#3)}
\def \prd#1#2#3{Phys. Rev. D {\bf#1}, #2 (#3)}
\def \prl#1#2#3{Phys. Rev. Lett. {\bf#1}, #2 (#3)}
\def \prp#1#2#3{Phys. Rep. {\bf#1}, #2 (#3)}
\def \ptp#1#2#3{Prog. Theor. Phys. {\bf#1}, #2 (#3)}
\def \rmp#1#2#3{Rev. Mod. Phys. {\bf#1}, #2 (#3)}
\def \rp#1{~~~~~\ldots\ldots{\rm rp~}{#1}~~~~~}
\def \si90{25th International Conference on High Energy Physics, Singapore,
Aug. 2-8, 1990}
\def \slc87{{\it Proceedings of the Salt Lake City Meeting} (Division of
Particles and Fields, American Physical Society, Salt Lake City, Utah, 1987),
ed. by C. DeTar and J. S. Ball (World Scientific, Singapore, 1987)}
\def \slac89{{\it Proceedings of the XIVth International Symposium on
Lepton and Photon Interactions,} Stanford, California, 1989, edited by M.
Riordan (World Scientific, Singapore, 1990)}
\def \smass82{{\it Proceedings of the 1982 DPF Summer Study on Elementary
Particle Physics and Future Facilities}, Snowmass, Colorado, edited by R.
Donaldson, R. Gustafson, and F. Paige (World Scientific, Singapore, 1982)}
\def \smass90{{\it Research Directions for the Decade} (Proceedings of the
1990 Summer Study on High Energy Physics, June 25--July 13, Snowmass,
Colorado),
edited by E. L. Berger (World Scientific, Singapore, 1992)}
\def \tasi90{{\it Testing the Standard Model} (Proceedings of the 1990
Theoretical Advanced Study Institute in Elementary Particle Physics, Boulder,
Colorado, 3--27 June, 1990), edited by M. Cveti\v{c} and P. Langacker
(World Scientific, Singapore, 1991)}
\def \yaf#1#2#3#4{Yad. Fiz. {\bf#1}, #2 (#3) [Sov. J. Nucl. Phys. {\bf #1},
#4 (#3)]}
\def \zhetf#1#2#3#4#5#6{Zh. Eksp. Teor. Fiz. {\bf #1}, #2 (#3) [Sov. Phys. -
JETP {\bf #4}, #5 (#6)]}
\def \zpc#1#2#3{Zeit. Phys. C {\bf#1}, #2 (#3)}

\begin{flushright}
TECHNION-PH-93-32 \\
EFI 93-41 \\
hep-ph/9308371 \\
September 1993 \\
\end{flushright}

\bigskip
\medskip
\begin{center}
\large
{\bf Identification of Neutral ${\bf B}$ Mesons using Correlated Hadrons}

\bigskip
\medskip

\normalsize
{\it Michael Gronau} \\

\medskip
{\it Department of Physics, Technion-Israel Institute of Technology \\
Technion City, 32000 Haifa, Israel}

\bigskip
and
\bigskip

{\it Jonathan L. Rosner \\
\medskip

Enrico Fermi Institute and Department of Physics \\
University of Chicago, Chicago, Illinois 60637 } \\

\bigskip
\bigskip
{\bf ABSTRACT}

\end{center}

\begin{quote}

The identification of the flavor of a neutral $B$ meson can make use of hadrons
produced nearby in phase space. Examples include the decay of ``$B^{**}$''
resonances or the production of hadrons as a result of the fragmentation
process. Some aspects of this method are discussed, including time-dependent
effects in neutral $B$ decays to flavor states, to eigenstates of CP and to
other states, and the effects of possible coherence between $B^0$ and
$\overline{B}^0$ in the initial state.  We study the behavior of the leading
hadrons in $b$-quark jets and the expected properties of $B^{**}$ resonances.
These are extrapolated from the corresponding $D^{**}$ resonances, of whose
properties we suggest further studies.
\end{quote}

\newpage

\centerline{\bf I.  INTRODUCTION}

\bigskip

For thirty years, the only system in which CP violation has been observed is
that of neutral kaons, for which several possible explanations exist \cite{WW}.
The most popular, that of phases in the Cabibbo-Kobayashi-Maskawa (CKM) matrix
\cite{CKM}, can be tested using systems of $B$ mesons \cite{Brefs}.

Asymmetries in rates of decays of neutral $B$ mesons to CP eigenstates such as
$J/\psi~K_S$ and $\pi^+ \pi^-$ are particularly easy to interpret in terms of
fundamental phases in the CKM matrix. However, the flavor of the decaying meson
[$B^0 ( = \bar{b} d) $ or $\overline{B}^0 (= b \bar{d})$] must be identified at
some reference time which is to be compared with the time of decay.

One suggestion for tagging the flavor of a produced neutral $B$ meson
\cite{GNR} is to study its correlation with pions produced nearby in phase
space. This method has been used to identify neutral $D$ mesons though the
decays of charged $D^*$ resonances \cite{Dtag}. The $D^{* \pm}$ resonance, with
spin\--parity $J^P = 1^-$, lies just above threshold for this decay, giving
rise to a characteristically soft pion. The corresponding $1^- B^*$ lies only
about $46$ MeV above the $B$, so $B^* \to \pi B$ is forbidden, but there are
positive\--parity resonances with $J^P = 0^+ , ~ 1^+$, and $2^+$ and masses
below about 5.8 ${\rm GeV}/c^2$ that are expected to couple to $\pi B$ and/or
$\pi B^*$. Even if these ``$B^{**}$'' resonances cannot  all be identified, one
expects some $\pi B^{(*)}$ correlations as a result of the fragmentation
process, as illustrated in Fig.~1.

In the present paper we expand upon the method proposed in Ref.~\cite{GNR}. The
number of $B$ mesons in modes such as $J/\psi ~K$ and $J/\psi ~K^*$
reconstructed by the CDF \cite{CDFB} and various LEP \cite{LEPB} collaborations
is large enough that the correlations of these mesons with pions nearby in
phase space are already under investigation \cite{CDFPC}. Our purpose is to
provide guidance for such studies.  As a corollary, we have found a general
description of initial ``tagged'' states of neutral $B$ mesons which may be of
use in any study of CP-violating decay asymmetries.

In Section II we generalize the approach of Ref.~\cite{GNR} to time-dependent
decays. In view of the considerable precision in $B$ lifetime measurements
which has already been achieved by the CDF and LEP groups, the study of
time-dependent effects may not be too far in the future.  Indeed, the
first results of such studies have already been presented by the ALEPH
\cn~\cite{ALtime}.

The measurement of time dependences in decays can shed light on the question,
addressed only briefly in Ref.~\cite{GNR}, of whether interferences between
$B^0$ and $\overline{B}^0$ produced in conjunction with a pion of a given
charge can ever occur. We assumed in Ref.~\cite{GNR}, and shall assume in the
present study, that in high energy $e^+ e^-$ collisions and in a hadronic
reaction $B^0$ and $\overline{B}^0$ are always incoherent with respect to one
another. In Sec.~III we show how to test this hypothesis by also allowing a
coherent or partially coherent admixture of $B^0$ and $\overline{B}^0$.  This
method has applications to any study of neutral $B$ mesons, as we have pointed
out in a shorter communication \cite{GRPRL}.

We mention several aspects of tagging $B$'s in Section IV, drawing attention
to methods using hadrons other than pions \cite{Ktag,ptag} and stressing the
importance of corresponding studies using charmed mesons. We also discuss the
question of whether explicit $B^{**}$ resonances are needed in order for the
method to succeed.

Section V is devoted to some general remarks on resonances which can decay to a
tagging hadron and a neutral $B$ or $B^*$. We refer the reader to
Refs.~\cite{Bres} and \cite{FP} for recent more detailed discussions of
properties of some of these resonances. As in Section IV, we stress the
importance of corresponding studies using charmed mesons. We conclude in
Section VI.

\bigskip
\centerline{\bf II.  TIME DEPENDENCES}

\bigskip

\noindent
A.  {\bf Identification of $B^0 - \overline{B}^0$ oscillations}
\bigskip

The study of same-sign lepton production in the reaction $e^+ + e^- \to
B^0 + \overline{B}^0$ led to the conclusion that the neutral $B$ meson
underwent significant mixing with its antiparticle \cite{oldmix}. The current
estimate of the mixing parameter, averaged over ARGUS and CLEO data, is
\cite{newmix} $\Delta m /\Gamma = 0.66 \pm 0.10$.

Explicit $B^0 - \overline{B}^0$ oscillations have been identified in
high\--energy $e^+ e^-$ collisions at LEP by the ALEPH collaboration
\cite{ALtime}. In the reaction $e^+ e^- \to Z^0 \to b \bar{b}$, the flavor of a
produced neutral $B$ meson is tagged by means of the semileptonic decay of the
$b$ quark not incorporated into this meson.

In Ref.~\cite{GNR} we defined the relative rates of production of $B^0$ and
$\overline{B}^0$ mesons in low\--mass combinations with charged pions to be
\begin{equation}
\begin{array}{c c c}
N (\overline{B}^0 \pi^- ) \equiv P_1 ~,& & N (\overline{B}^0 \pi^+ )
\equiv P_2 \\
& & \\
N (B^0 \pi^+) \equiv P_3 ~,& & N (B^0 \pi^- )
\equiv P_4  ~.\\
\end{array}
\end{equation}
For $e^+e^- \to Z^0 \to b \bar{b}$ and for $\bar p p \to B \pi + \ldots$,
charge conjugation symmetry implies $P_3 = P_1 , ~ P_4 = P_2$. Let us imagine
that a neutral $B$ decays to a state of identifiable flavor, e.g.,
\begin{equation}
B^0 \to J/ \psi ~ K^{*0} ~, ~~~
\overline{B}^0 \to J/\psi ~ \overline{K}^{*0} ~~,
\end{equation}
with the flavor of the neutral $K^*$ identified by the decay $K^{*0} \to K^+
\pi^-$ or $\overline{K}^{*0} \to K^- \pi^+$. Let us denote a ``right\--sign''
combination $R$ as $\overline{B}^0 \pi^-$ or $B^0 \pi^+$, and a
``wrong\--sign'' combination $W$ as $\overline{B}^0 \pi^+$ or $B^0 \pi^-$. Then
as a function of proper decay time, the relative numbers of right\--sign and
wrong sign combinations are:
\begin{equation}
R(t) = e^{-\Gamma t} \left [ P_1 \cos^2 \left ( \frac{\Delta m t}{2} \right ) +
P_2 \sin^2 \left ( \frac{\Delta m t}{2} \right ) \right ] ~,
\end{equation}
\begin{equation}
W(t) = e^{-\Gamma t} \left [ P_1 \sin^2 \left ( \frac{\Delta m t}{2} \right ) +
P_2 \cos^2 \left ( \frac{\Delta m t}{2} \right ) \right ] ~,
\end{equation}
so that the time-dependent asymmetry is given by
\begin{equation}
\label{TDA}
\frac{R(t) - W(t)}{R(t) + W(t)} =
\frac{P_1 - P_2}{P_1 + P_2} \cos (\Delta m t) ~~~.
\end{equation}
(Here we have ignored very small CP-violating effects in the decays in
question.) The corresponding time\--integrated asymmetry is
\begin{equation}
\label{TIA}
\frac{\int [R(t) - W(t)]dt}{\int [R(t) + W(t) ] dt} =
\frac{P_1 - P_2}{P_1 + P_2} ~ \frac{1}{1 + x_d^2} ~~,
\end{equation}
where $x_d \equiv (\Delta m / \Gamma )_{B^0}$. The factor $[1 + x_d^2]^{-1}$ is
about $2/3$. Thus, since time\--dependent information is available anyway in
extraction of the $B$ signals from many experiments, Eq.~(\ref{TDA}) may
provide information on $(P_1 - P_2)/(P_1 + P_2)$ which is at least as
statistically compelling as the time\--integrated asymmetry (\ref{TIA}).

Of course, in the decays of charged $B$'s, e.g. to $J/\psi ~K^{\pm}$, no $B -
\overline{B}$ oscillations will occur, and one will measure just the dilution
factor $(P_1 - P_2)/(P_1 + P_2)$ when forming the corresponding right sign --
wrong sign asymmetry. Here, with $r \equiv B^- \pi^+$ or $B^+ \pi^-$ and $w
\equiv B^- \pi^-$ or $B^+ \pi^+$, one has
\begin{equation}
(r-w)/(r+w) = (P_1 - P_2)/(P_1 + P_2) ~~~.
\end{equation}
The comparison of this result with (\ref{TDA}) or (\ref{TIA}) will form a
useful test of isospin independence of the production process. Such
independence is frequently but not universally expected to occur \cite{GNR}. A
specific case in which it could be violated would be if a meson is produced by
fragmentation of a proton into a $b$\--flavored baryon and a meson containing a
$\bar{b}$, as shown in Fig.~2. Since the proton has more $u$ than $d$ valence
quarks, one might expect more $B^+$  than $B^0$ mesons in such a process.
\bigskip

\noindent
B.  {\bf Time-dependent CP-violating asymmetries}
\bigskip

The time-integrated asymmetry for decays of states of identified flavor
at $t=0$ into a CP eigenstate $f$ may be defined as
\begin{equation}
A(f) \equiv
\frac{\Gamma (B_{t=0}^0 \to f) - \Gamma (\overline{B}_{t=0}^0 \to f)}{\Gamma
 (B_{t=0}^0 \to f) + \Gamma (\overline{B}_{t=0}^0 \to f)} ~~~~.
\end{equation}
For $f = J/\psi ~K_S$, one has
\begin{equation}
A(J/\psi ~ K_S ) = -
\frac{x_d}{1 + x_d^2} \sin (2 \beta) ~~~,
\end{equation}
where $\beta$ is an angle in the triangle expressing unitarity of the
Cabibbo\--Kobayashi\--Maskawa (CKM) matrix \cite{Brefs}. In cases in which
$B^0$ and $\overline{B}^0$ production cross sections are equal, we derived in
\cite{GNR} the simple relation
\begin{equation}
A_{\rm obs}(f, \pi) =
\frac{P_1 - P_2}{P_1 + P_2} A(f) ~~~,
\end{equation}
where the observed asymmetry is defined in terms of the charged pion.
The corresponding time-{\em dependent} rates for states which were initially
$B^0$ or $\overline{B}^0$ at $t=0$, aside from common overall factors, are
\begin{equation}
\label{TDG}
\Gamma (t) \equiv
\frac{d \Gamma }{dt}(B_{t=0}^0 \to f) = e^{- \Gamma t} (1- \sin 2 \beta \sin
\Delta mt)
\end{equation}
\begin{equation}
\label{TDGbar}
\overline{\Gamma} (t) \equiv
\frac{d \Gamma}{dt} (\overline{B}_{t=0}^0 \to f) = e^{- \Gamma t}
(1 + \sin 2 \beta \sin
\Delta mt)
\end{equation}
so the time-dependent asymmetry is
\begin{equation}
\label{TDGA}
A(t) =
\frac{\Gamma (t) - \overline{\Gamma}(t)}{\Gamma (t) + \overline{\Gamma}(t)}
= - \sin 2 \beta \sin (\Delta m t)
\end{equation}
The observed asymmetry associated with pions of opposite charge will be diluted
by a common factor $(P_1 - P_2)/(P_1 + P_2)$, as in the time-integrated case.

It may be necessary to take into account the explicit time-dependences
(\ref{TDG}) and (\ref{TDGbar}) when discussing efficiencies for detecting $B$
mesons. Typically such efficiencies vary as a function of proper decay
lifetime.

\bigskip
\noindent
{\bf C. States which are not CP eigenstates}
\bigskip

Angles of the unitarity triangle can also be determined from neutral $B$ decays
to states $f$ which are not CP eigenstates. This is feasible when both a $B^0$
and a $\overline B^0$ can decay to a final state $f$ which appears in only one
partial wave, provided that a single weak CKM phase dominates each of the
corresponding decay amplitudes. Two interesting examples are \cite{nonCP}
$B^0_d\to \rho^-\pi^+$, for which one must neglect the penguin amplitude, and
$B^0_s\to D^-_s K^+$, where a single amplitude is known to contribute in the
standard model.  In both these cases, in contrast to the form (\ref{TDGA}), one
sees both sines and cosines of $\Delta m t$.

Let us comment in passing on the specific signatures of the decays $B_s
{\rm~or}~ \overline{B}_s$ $\to D_s^+ K^-$.  The graphs contributing to this
process are shown in Fig.~3.  A good mode for detecting the $D_s^+$ is via its
$\phi \pi^+$ decay, where $\phi \to K^+ K^-$.  The final state of the strange
$B$ then contains three charged kaons and one charged pion coming from the
secondary vertex.  As we shall discuss in more detail below, the initial flavor
of a strange $B$ is to be determined on a statistical basis by means of
correlations with a charged kaon \cite{Ktag}.  This kaon comes from the primary
production vertex.  It would be highly desirable to invent a trigger for
hadronically produced events containing four charged kaons to increase the
sensitivity for such events.

We wish to demonstrate how our tagging method can be applied to a general case
of a non-CP eigenstate. The time-dependent rates for states which were $B^0$ or
$\overline B^0$ at $t=0$ and decay at time $t$ to the state $f$ or its
CP-conjugate $\overline f$ are given by \cite{GR89}:
\begin{eqnarray}
\Gamma_f(t) & = & e^{-\Gamma t} [ |A|^2\cos^2({\Delta m t\over 2}) +
|\bar A|^2\sin^2({\Delta m t\over 2}) \nonumber \\
{}~& + & |A\bar A|\sin(\delta + \phi_M +
\phi_D)\sin(\Delta m t) ]~~~, \nonumber \\
\overline{\Gamma}_f(t) & = & e^{-\Gamma t} [ |\bar A|^2\cos^2({\Delta m t
\over 2}) + |A|^2\sin^2({\Delta m t\over 2}) \nonumber \\
{}~& - & |A\bar A|\sin(\delta +
\phi_M + \phi_D)\sin(\Delta m t) ]~~~, \nonumber \\
\Gamma_{\bar f}(t) & = & e^{-\Gamma t} [ |\bar A|^2\cos^2({\Delta
m t\over 2}) + |A|^2\sin^2({\Delta m t\over 2}) \nonumber \\
{}~& - & |A\bar A|\sin(\delta -
\phi_M - \phi_D)\sin(\Delta m t) ]~~~, \nonumber \\
\overline{\Gamma}_{\bar f}(t) & = & e^{-\Gamma t} [ |A|^2\cos^2({\Delta m
t \over 2}) + |\bar A|^2\sin^2({\Delta m t\over 2}) \nonumber \\
{}~& + & |A\bar A|\sin(\delta
- \phi_M - \phi_D)\sin(\Delta m t) ]~~~.
\end{eqnarray}
Here $|A|$ and $|\bar A|$ are the magnitudes of the decay amplitudes of $B^0$
and $\overline B^0$ to $f$, $\delta$ and $\phi_D$ are the strong and weak
phase-differences between these amplitudes, and $\phi_M$ is the phase of
$B-\overline B$ mixing. The corresponding time-dependent rates for states $f$
or $\bar f$ in conjunction with pions of positive or negative charges are then:
\begin{eqnarray}
\Gamma_{f\pi^{\pm}} (t) = (1/2) e^{-\Gamma t} \{ |A|^2+
|\bar A|^2 \pm [P_1-P_2][ (|A|^2-|\bar A|^2)\cos(\Delta m t) \nonumber \\
+2|A\bar A|\sin(\delta + \phi_M + \phi_D)\sin(\Delta m t) ] \}~~~, \nonumber \\
\Gamma_{\bar f\pi^{\pm}} (t) = (1/2) e^{-\Gamma t} \{ |A|^2+
|\bar A|^2 \mp [P_1-P_2][ (|A|^2-|\bar A|^2)\cos(\Delta m t) \nonumber \\
+2|A\bar A|\sin(\delta - \phi_M - \phi_D)\sin(\Delta m t)] \}~~~,
\end{eqnarray}
where we have taken $P_1 + P_2 = 1$.

These four rates depend on four unknown quantities, $|A|,~|\bar A|,~
\sin(\delta + \phi_M + \phi_D)$ and $\sin(\delta - \phi_M - \phi_D)$.
Measurement of the rates allows a determination of the weak CKM phase $\phi_M +
\phi_D$, apart from a two-fold ambiguity \cite{nonCP}. In the two cases
$B^0_d\to\rho^-\pi^+$ and $B^0_s\to D^-_s K^+$ this phase obtains the values
$2\alpha$ and $\gamma$, respectively.

\bigskip
\centerline{\bf III.  THE QUESTION OF COHERENCE}
\bigskip

The search for CP-violating asymmetries in decays of neutral $B$ mesons
produced at the $\Upsilon(4S)$ resonance involves correlations between the
particle whose decay is studied and the particle whose decay serves to ``tag''
the flavor of its partner.  Here, coherence between a $B^0$ and a
$\overline{B}^0$ is crucial.  There have been several studies
\cite{Brefs,Bspin} of such coherence, both at the $\Upsilon(4S)$, where a $B^0
\overline{B}^0$ pair is in a state with $C = -1$, and in configurations where
an extra photon has been produced leading to $C = + 1$ for the $B^0
\overline{B}^0$ pair.

In a hadronic production environment or in high energy $e^+ e^-$ reactions, it
is much less likely to find coherence between a $B$ and $\overline B$, since
they are usually separated in rapidity by many intermediate hadrons.  The
absence of coherence was an assumption which was made not only in
Ref.~\cite{GNR} but which appears in many other treatments of hadronic
production \cite{CF}.  It seems prudent to test for such coherence directly.
We have found that such a test is possible, and describe it briefly in the
present section.  We have reported these results in more detail in
Ref.~\cite{GRPRL}.

We denote particle and antiparticle basis states by spinors with spin up
and spin down in an abstract ``quasispin'' space \cite{Bspin,CPrevs}:

\begin{equation}
| B^0 \rangle = \left[ \begin{array}{c} 1 \\ 0 \\ \end{array} \right]~~~,~~
| \overline B^0 \rangle = \left[ \begin{array}{c} 0 \\ 1 \\ \end{array}
\right]~~~.
\end{equation}
A density matrix $\rho$ allows one to discuss incoherent and coherent states in
a unified manner.  The most general such $2 \times 2$ matrix has the form $\rho
= (1 +  {\bf Q \cdot \sigma})/2$, where ${\bf Q}$ is a vector describing
polarization in quasispin space, satisfying ${\bf Q}^2 \leq 1$, and
$\sigma_i~(i = 1,~2,~3)$ are the Pauli matrices.

A pure state corresponds to a linear combination of $B^0$ and
$\overline{B}^0$ with arbitrary complex coefficients, whose sum of absolute
squares equals unity.  Such a state can be denoted by a density matrix with $Q
\equiv |{\bf Q}| = 1$. An arbitrary incoherent combination of $B^0$ and
$\overline{B}^0$ with relative probabilities $P_1$ and $P_2 = 1 - P_1$
corresponds to a diagonal density matrix with $Q_1 = Q_2 = 0,~Q_3 = 2P_1 - 1$.
This is the case we considered to hold in Ref.~\cite{GNR}. As a special case of
either example, one describes the density matrices for initial $B^0$ and
$\overline{B}^0$ by diag(1,0) and diag(0,1), respectively.

The probability for a transition from an initial state denoted by the density
matrix $\rho_i$ to a final state denoted by $\rho_f$ is then $ I(f) = {\rm
Tr}~(\rho_i {T}^{\dag} \rho_f T)$, where $T$ is the operator which time-evolves
the state from $i$ to $f$.  Here $\rho_f$ can denote an arbitrary coherent
superposition of $B^0$ and $\overline{B}^0$ at time $t$, or can also take
account of the {\it decay} of this superposition.

To discuss mass eigenstates, which have simple time-evolution properties, we
neglect differences between their lifetimes \cite{Brefs}, and denote them by
$B_L$ (``light'') and $B_H$ (``heavy''):
\begin{equation}
|B_L \rangle = (|B^0 \rangle +  |\overline{B}^0 \rangle)/\sqrt{2}~~~,~~
|B_H \rangle = (|B^0 \rangle -  |\overline{B}^0 \rangle)/\sqrt{2} ~~~.
\end{equation}
We adopt a convention in which the $B^0 - \overline{B}^0$ mixing amplitude is
real, and use the fact \cite{Brefs} that the mixing amplitudes are of
approximate magnitude $1/\sqrt{2}$. This differs from a more standard
convention by a phase which we take into account when calculating amplitudes
for decays of $b$ quarks. The transformation between flavor eigenstates and
mass eigenstates is then implemented by the unitary matrix $U = (\sigma_1 +
\sigma_3)/\sqrt{2}$.  The matrix describing the time evolution in
the mass eigenstate basis is

\begin{equation}
e^{-i M_D t} \equiv e^{- \Gamma t/2}
{\rm diag}(e^{-i m_L t}, e^{-i m_H t})
= e^{- \Gamma t/2} e^{- i \bar m t} e^{i \sigma_3 \Delta m t /2}~~~,
\end{equation}
where $\bar m \equiv (m_H + m_L)/2,~\Delta m \equiv m_H -m_L$.
Thus the time evolution operator $T$ in the $B^0, \overline{B}^0$ basis is
just $T = {U}^{\dag} e^{-i M_D t} U = e^{- \Gamma t/2}
e^{-i \bar m t} e^{i \sigma_1 \Delta m t/2}$.

The trace for the transition probability $I(f)$ can be computed by applying the
matrices $U$ and ${U}^{\dag}$ to the initial and final density matrices.
Defining $\rho' \equiv U \rho {U}^{\dag}$, we find that the effect of $U$ is to
rotate ${\bf Q}$ into ${\bf Q}'$, where
\begin{equation}
Q_1' = Q_3~~~,~~Q_2' = - Q_2~~~,~~Q_3' = Q_1 ~~~.
\end{equation}
The transition probability can now be written in terms of traces as
\begin{equation}
I(f) = {\rm Tr}~ (\rho_i' e^{i M^*_D t} \rho_f' e^{-i M_D t} )~~~.
\label{treq}
\end{equation}

The states $B^0$ and $\overline{B}^0$ at the time of decay $t$ may be
identified by their decays to states of identifiable flavor, e.g., $B^0 \to
J/\psi K^{*0}$, with $K^{*0} \to K^+ \pi^-$.  We assume that a single weak
subprocess contributes to the decay, which is an excellent approximation for
these final states \cite{Brefs}.

With the convention in which the mixing amplitudes in the neutral
$B$ mass eigenstates are real, the weak decay amplitudes for $B^0 \to J/\psi
K^{*0}$ and $\overline{B}^0 \to J/\psi \overline{K}^{*0}$ may be denoted $A
e^{-i \beta}$ and $Ae^{i \beta}$, respectively, where $\beta = {\rm Arg}~(-
V_{cb}^* V_{cd}/V_{tb}^* V_{td})$, and $V_{ij}$ are elements of the
Cabibbo-Kobayashi-Maskawa matrix specifying the charge-changing weak couplings
of quarks.  We find

\begin{equation}
I \left( \begin{array}{c} B^0 \\ \overline{B}^0 \\ \end{array} \right) =
\frac{1}{2} |A|^2 e^{-\Gamma t} \left[ 1 \pm Q_{\perp}' \cos (\Delta m t +
\delta) \right]~~~,
\label{flavor1}
\end{equation}

\begin{equation}
Q_1' \equiv Q_{\perp}' \cos \delta~~~,~~
Q_2' \equiv Q_{\perp}' \sin \delta~~~.
\label{defs}
\end{equation}

When the initial state is an incoherent mixture of $B^0$ and $\overline{B}^0$
with relative probabilities $P_1$ and $P_2 = 1 - P_1$, respectively, one sets
$Q'_\perp = 2 P_1 - 1$ and $\delta = 0$ in the above expression, recovering the
results of Sec.~II A.

We consider a charge-symmetric production process in which an arbitrary
combination of neutral $B^0$ and $\overline B^0$ is produced, with an
additional particle of specific charge (such as a charged lepton or pion)
bearing some specific kinematic relation to it. We wish to determine the
relation of this process to the one in which the charge of the tagging particle
is the opposite.

Under the phase convention we have chosen for $B^0$ and $\overline{B}^0$, if we
take $CP|B_L \rangle = |B_L \rangle$ and $CP|B_H \rangle = -|B_H \rangle $, the
charge conjugation operation has the phase
\begin{equation}
C |B^0 \rangle = - |\overline{B}^0 \rangle ~~~;~~
C |\overline{B}^0 \rangle = - |B^0 \rangle ~~~.
\end{equation}
Under charge conjugation, the first and second rows and columns of the density
matrix $\rho$ are interchanged, so that $Q_1 \to Q_1,~Q_2 \to -Q_2,~Q_3 \to -
Q_3$, or $Q'_1 \to -Q'_1,~Q'_2 \to -Q'_2,~Q'_3 \to Q'_3$. Therefore, the decay
rates $\bar I(f)$ for states tagged with antiparticles are given in terms of
those $I(f)$ for states tagged with particles by
\begin{equation}
\bar I(f;~Q'_1,~Q'_2,~Q'_3) = I(f;~-Q'_1,~-Q'_2,~Q'_3)~~~.
\end{equation}

For a final state identified as a $B^0$ by its decay to $J/\psi K^{*0}$, the
time-dependent asymmetry is
\begin{equation}
A(J/\psi K^{*0}) \equiv \frac{I(J/\psi K^{*0}) - \bar I(J/\psi K^{*0})}
{I(J/\psi K^{*0}) + \bar I(J/\psi K^{*0})}
= Q'_\perp \cos(\Delta m t + \delta)~~~.
\end{equation}
As a consequence of the assumed charge symmetry of the production process, one
has $\bar I(J/\psi K^{*0}) = I(J/\psi \overline{K}^{*0})$ and $\bar I(J/\psi
\overline{K}^{*0}) = I(J/\psi K^{*0})$.

We can then measure the components $Q'_\perp$ and $\delta$ using decays to
flavor eigenstates. (The ALEPH Collaboration \cite{ALtime} has measured
a time-dependent asymmetry of the above form, fitting it under the assumption
$\delta = 0$.)

A measurement of $Q'_3$ for neutral nonstrange $B$ mesons can be performed by
utilizing their decays to the specific CP eigenstate $J/\psi~K_S$.  In our
phase convention, the amplitudes for $B^0$ and $\overline{B}^0$ to decay into
$J/\psi K_S$ are $A' e^{-i \beta}/\sqrt{2}$ and $-A' e^{i \beta}/\sqrt{2}$.
Here we have taken into account the intrinsic negative CP of the $J/\psi K_S$
state, and neglected the small CP violation in the kaon system. The density
matrix for the final state is then
\begin{equation}
\rho_{J/\psi K_S} = \frac{1}{2} |A'|^2 \left[ e^{-i \beta}~-e^{i \beta}
\right]~ \left[ \begin{array}{c} e^{i \beta} \\ - e^{-i \beta} \end{array}
\right] = \frac{1}{2} |A'|^2 \left[ \begin{array}{c c} 1 & -e^{2 i \beta} \\
-e^{-2 i \beta} & 1 \\ \end{array} \right]~~~,
\end{equation}
with off-diagonal terms in $\rho$ changed in sign for $J/\psi K_L$.
The expressions for the decay rates for states prepared with particle and
antiparticle tags are
\begin{eqnarray}
I \left( J/\psi \begin{array}{c} K_L \\ K_S \\ \end{array}\right) & = &
\frac{1}{2} |A'|^2 e^{-\Gamma t} \left\{ 1 \pm [Q'_3 \cos 2 \beta + Q'_\perp
\sin 2 \beta~ \sin (\Delta m t + \delta )] \right\}, \nonumber \\
\bar I \left( J/\psi \begin{array}{c}K_L \\ K_S \\ \end{array}\right) & = &
\frac{1}{2} |A'|^2 e^{-\Gamma t} \left\{ 1 \pm [Q'_3 \cos 2 \beta - Q'_\perp
\sin 2 \beta~ \sin (\Delta m t + \delta )] \right\}. \nonumber \\
\end{eqnarray}
As in the case of decays to the flavor eigenstates, the time-dependent term has
a phase shift $\delta$ and a modulation amplitude $Q'_\perp$.
The decay asymmetry for the $J/\psi K_S$ final state is
\begin{equation}
A(J/\psi K_S) \equiv \frac{I(J/\psi K_S) - \bar I(J/\psi K_S)}
{I(J/\psi K_S) + \bar I(J/\psi K_S)}
= \frac{- Q'_\perp ~\sin 2 \beta ~\sin(\Delta m t + \delta)}
{1 - Q'_3 \cos 2 \beta}~~~.
\end{equation}

The component $Q'_3$ (which appears even in the absence of CP violation)
is a necessary ingredient in the discussion of possible coherence. It is this
component that leads to correlations between $K_S$ and $K_L$ produced in
$\phi$ decay, as discussed in Ref.~\cite{Lip68}.  In order to learn its value,
we measure the rate for $J/\psi K_S$ production (summing over particle and
antiparticle tags, so that the time-dependent terms cancel). We compare this
with the corresponding sum of rates (which also has no time-dependence, and is
independent of $Q'_3$) for production of a flavor eigenstate $K^0$.

To measure the $K^0$ production rate, there are a couple of possibilities.  If
the rate of production of charged and neutral $B$'s is equal, as is expected
for high-energy $e^+ e^-$ collisions \cite{GNR}, one can use the observed rate
for $J/\psi K^+$ production, making use of the fact that the decays of $B$
mesons to $J/\psi K$ involve the quark subprocess $b \to c \bar c s$, which
conserves isospin \cite{LS}. Or, one can measure the $B^0/B^+$ production ratio
via the observed $K^{*0}/K^{*+}$ ratio, and infer the $K^0$ rate from the
observed $K^+$ rate.  In short, the ratio $|A'/A|^2$ is measurable.

Once we learn the relative normalization of rates for decays to flavor
eigenstates and CP eigenstates, we can determine the magnitude of the term
$Q_3' \cos 2 \beta$, and then use the asymmetry in $J/\psi K_S$ decay to
measure $ \sin 2 \beta$.  With the possibility of a discrete ambiguity
(unlikely for known ranges of CKM parameters), we then obtain $\cos 2 \beta$,
thereby finding $Q'_3$ itself.

The corresponding decay asymmetry for the $\pi^+ \pi^-$ final state is easily
calculated.  We neglect penguin effects \cite{pen}, which can be
dealt with by studying the $2 \pi^0$ final state.  With our phase convention
for $b$ quarks, the result can be obtained by the substitution $\beta \to -
\alpha$ in the corresponding result for the $J/\psi K_L$ final state, where
$\alpha = {\rm Arg}~(-V_{tb}^* V_{td}/V_{ub}^* V_{ud})$. We find
\begin{equation}
A(\pi^+ \pi^-) \equiv \frac{I(\pi^+ \pi^-) - \bar I(\pi^+ \pi^-)}
{I(\pi^+ \pi^-) + \bar I(\pi^+ \pi^-)}
= \frac{- Q'_\perp ~\sin 2 \alpha ~\sin(\Delta m t + \delta)}
{1 + Q'_3 \cos 2 \alpha}~~~.
\end{equation}
Since we have already measured all components of ${\bf Q}'$ and the phase
$\delta$, this result can be used to extract $\alpha$.

\bigskip

\centerline{\bf IV.  FURTHER REMARKS ON TAGGING}

\bigskip

\noindent
A.  {\bf Resonances vs. more general correlations}
\bigskip

The fragmentation diagrams shown in Fig.~1 indicate that a correlation between
the charge of the leading pion and the flavor of the neutral $B$ is possible
{\em whether or not} that pion resonates with the $B$ (or its parent $B^*$,
decaying to $B \gamma$).  Very recently this correlation was calculated for LEP
energies \cite{GG} using a soft fragmentation version of JETSET 7.3.  It was
found that the correlation factor $[N(B^0 \pi^+) - N(B^0 \pi^-)]/ [N(B^0 \pi^+)
+ N(B^0 \pi^-)]$, for pions with the lowest $M(B \pi)$ value in each event,
increases from a value of 0.17 at $M(B \pi) = 5.5$ GeV/$c^2$ to the value of
0.27 at 5.8 GeV/$c^2$, and stays constant up to 6.2 GeV/$c^2$, where very small
rates are expected.  This estimate, which does not include resonance effects,
is quite encouraging, and should be tested experimentally.

It is an observed feature of hadron physics, however, that whenever a quark $q$
is contained in a meson $M_1$ and the same antiquark $\bar{q}$ is contained in
another meson $M_2$, the mesons $M_1$ and $M_2$ form their first resonance no
higher than several hundred MeV above threshold \cite{crf}. Moreover,
meson\--meson scattering in such ``non\--exotic'' channels (such as $\pi^+
\pi^-$ or $K^\pm \pi^\mp )$ is substantially stronger than in ``exotic''
channels like $\pi^\pm \pi^\pm$ or $K^\pm \pi^\pm$, {\em even at nonresonant
energies}.

The advantage of explicit $\pi B$ or $\pi B^*$ resonances, as stressed in
Ref.~\cite{GNR}, may be particularly great in eliminating combinatorial
backgrounds rather than in obtaining a correlation. This advantage is likely to
be most pronounced for narrow resonances. As calculated in \cite{Bres} and
mentioned in \cite{GNR} and below, we expect one of the $J^P = 1^+$ resonances
and the $J^P = 2^+$ resonance to be narrow, but the other $J^P = 1^+$ resonance
and the $J^P = 0^+$ resonance are likely to be considerably broader.

\bigskip
\noindent
B.  {\bf Tagging using hadrons other than pions}
\bigskip

For completeness, we wish to mention methods for tagging neutral nonstrange and
strange $B$ mesons which rely upon their correlations with kaons \cite{Ktag}
and protons \cite{ptag}. The corresponding fragmentation diagrams are shown in
Fig.~4 for correlations of kaons with nonstrange $B$'s, Fig.~5 for correlations
of kaons with strange $B$'s, and Fig.~6 for correlations of protons or
antiprotons with nonstrange $B$'s.

The common feature of all these methods is that the neutral $B$ meson contains
the same quark or antiquark as the corresponding antiquark or quark in the
tagging particle, and that this should serve to {\em uniquely specify} the
tagging particle. Thus, in Fig.~4, it would not be suitable to use a $K^0$ or
$\overline{K}^0$ as a tagging particle, since one would actually observe $K_S^0
\to \pi^+ \pi^-$.

Generalizations of the diagrams in Figs.~4-6 are easily made.

\bigskip

\noindent
C.  {\bf Lessons from correlations of hadrons with $D$ and $D^*$ mesons}
\bigskip

The presence of positive-parity ``$D^{**}$'' resonances has already been
established. We shall discuss their properties more explicitly in Sec.~V,
since they can provide valuable information about the corresponding $B^{**}$
resonances. Here we wish to note that the same sorts of correlations can be
studied for charmed mesons as for $B$ mesons. This type of study is not
essential for tagging neutral $D$ mesons themselves, since the $D$ decays
$D^{*+} \to \pi^+ D^0$ and $D^{*-} \to \pi^- D^0$ \cite{Dtag} are ideal for
that. The purpose of studying hadron\--$D$ or hadron\--$D^*$ correlations in
which the effective mass lies above the $D^*$ is to calibrate what sorts of
correlations one might expect for systems containing $b$ quarks.

As one example, consider the correlations of a charged kaon and a $D_s^{(*)}$
or $\overline{D}_s^{(*)}$, as shown in Fig.~7. These diagrams are identical to
those in Fig.~5 with the substitution of a charmed quark for a $b$ quark.
Moreover, both $B_s^{*0}$ and $D_s^{*+}$ decay via photon emission to $B_s^0$
and $D_s^+$, respectively. Thus, aside from the fact that we expect \cite{RW}
$M(B_s^{*0}) - M(B_s^0) = $ $M(B^{*0}) - M(B^0) \simeq 46$ MeV while we have
$M(D_s^{*+}) - M(D_s^+) \simeq 141$ MeV, the two systems should be very
similar. Differences could arise as a result of these different hyperfine
splittings if there are resonances very close to threshold; this possibility is
assessed in Sec.~V.

\bigskip
\centerline{\bf V.  RESONANCE LORE}

\bigskip

\noindent
A.  {\bf Positive-parity D mesons}
\bigskip

The bound states of a charmed quark $c$ with a light anti\--quark $\bar{q}$ in
an $L=1$ system have been discussed in many places, including Refs.~
\cite{DGG}, \cite{cmts}, and \cite{IW}. The understanding of such resonances
will help in anticipating the properties of the corresponding mesons involving
$b$ quarks.

The fine structure of the $L = 1~ c \bar{q}$ system is dominated by whether the
sum ${\bf L} + {\bf S_q} \equiv {\bf j}$ corresponds to $j = 1/2$ or $3/2$. The
states with $j = 1/2$ and their expected decay modes are:
\begin{equation}
J_{2j}^P = 0_1^+ : ~~ \to (D \pi)_{\ell =0} ~~,
\end{equation}
\begin{equation}
J_{2j}^P = 1_1^+ : ~~ \to (D^* \pi )_{\ell = 0}~~,
\end{equation}
Neither of these states has been observed yet. The states with $j = 3/2$ are
expected to be:
\begin{equation}
J_{2j}^P = 1_3^+ : ~~ \to (D^* \pi)_{\ell = 2} ~~,
\end{equation}
\begin{equation}
J_{2j}^P = 2_3^+ : ~~ \to (D \pi)_{\ell = 2} , (D^* \pi)_{\ell = 2} ~~.
\end{equation}
The states in these two pairs of equations are expected to be split by an
interaction whose strength depends on one inverse power of the heavy quark
mass.

Candidates for the $1_3^+$ and $2_3^+$ states exist \cite{ARGD,CLD,E687,E691}:
\begin{equation}
D^*(2420) \to D^* \pi~~~,
\end{equation}
\begin{equation}
D^* (2460) \to D \pi,~D^* \pi ~~~.
\end{equation}
The identification of the $2_3^+$ state is unique just on the basis of decay
modes. The identification of the $1_3^+$ state is supported by the small mass
splitting between the states and by the Dalitz plot distribution in the $D \pi
\pi$ final state. This distribution is consistent with the production of an
$\ell = 2 ~ D^* \pi$ final state \cite{CLD,E687}.

Adjusting the predictions of Ref.~\cite{DGG} to make the $1_3^+$ and $2_3^+$
states correspond to the observed ones, one then expects the $0_1^+$ and
$1_1^+$ states and to show up around 2.34 and 2.35 ${\rm GeV}/c^2$,
respectively. Other predictions for these states have been summarized in
Ref.~\cite{cmts}. A recent interesting suggestion \cite{BH} is that these
particles could be the parity doublets of the $0^- D$ and $1^- D^*$ mesons,
split from them by chiral symmetry breaking.

The failure to observe the $0_1^+$ and $1_1^+$ states up to now has usually
been ascribed to their ability to decay via $S$\--waves, and thus to be
extremely broad. It is important, nonetheless, to see if such states can be
identified, perhaps by comparison with exotic channels. Thus, for instance, to
search for the $0_1^+$ state one might compare $\pi^+ D^0$ (non\--exotic) and
$\pi^-D^0$ (exotic) channels, while to search for the $1_1^+$ state one might
compare $\pi^- D^{*+}$ (non\--exotic) and $\pi^+ D^{*+}$ (exotic) channels.

One also expects $0_1^+ , 1_1^+ , 1_{3}^+$ and $2_{3}^+$ {\em strange}
charmed mesons, about $100$ MeV above the corresponding nonstrange ones. (This
is about the observed splitting between the $D_s^+$ and the $D^+$, and between
the $D_s^{*+}$ and the $D^{*+}$.) A candidate for the $1_{3}^+$ strange
state has been seen \cite{Ds1}:
\begin{equation}
D_s^* (2536) \to D^* \overline{K} ~~~,
\end{equation}
The absence of a $D \overline{K}$ mode suggests that this is not the $2_{3}^+$
state.

\bigskip

\noindent
B.  {\bf Extrapolation to positive-parity B mesons}
\bigskip

A detailed study of the spectroscopy of $L=1 ~ b \bar{q}$ mesons has recently
been performed in Ref.~\cite{Bres}. Some earlier treatments are contained in
Ref.~\cite{masses}. Here we comment on those features which can be obtained
primarily from extrapolating the known or expected properties of the $L=1 ~ c
\bar{q}$ mesons.

The fine-structure splitting between the states $1_{3}^+$ and $2_{3}^+$ scales
as $1/m_Q$, where $Q$ is the heavy quark. Thus, we expect the corresponding $b
\bar{q}$ states to be split by $m_c / m_b \simeq 1/3$ times the splitting in
the charm system, or about $13$ MeV. Now, the spin\--weighted average of the
charmed $1_3^+$ and $2_3^+$ masses is about $2445~{\rm MeV}/c^2$, which lies
about $470~{\rm MeV}/c^2$ above the spin-weighted average of the $D$ and $D^*$
masses. Thus, if the dynamics of the $c \bar{q}$ and $b \bar{q}$ systems are
similar, we expect the spin\--weighted average of nonstrange $1_{3}^+$ and
$2_{3}^+ ~ b \bar{q}$ states to lie about $470 ~ {\rm MeV}/c^2$ above $[3M(B^*)
+ M(B)]/4 \simeq 5313$ ${\rm MeV}/c^2$, or at $5783 ~ {\rm MeV}/c^2$. (Taking
account of the slightly greater binding energy of the $b \bar q$ system, the
authors of Ref.~\cite{Bres} find this value to be 20 MeV$/c^2$ lower.)

The $(1_{3}^+ , ~ 2_{3}^+)$ states should then lie at $(5775,~5788)$ ${\rm
MeV}/c^2$ (or (5755,~5767) ${\rm MeV}/c^2$ in the estimate of
Ref.~\cite{Bres}). The $(0_{1}^+ , ~ 1_{1}^+)$ states should lie about $100$
MeV lower. For the corresponding strange states, one should add about $100$
MeV. (This appears to be true in comparing the $B^0$ with the recently observed
$B_s^0$ \cite{Bs}, and in comparing nonstrange and strange $J^P = 1^+$ charmed
mesons.)  We summarize these expectations in Table I.

\begin{center}
Table I.  Expected properties of $L=1 ~ b \bar{q}$ states.

\medskip
\begin{tabular}{c | c l | c l }
\multicolumn{3}{c}{$\bar{q} = \bar{u}$ or $ \bar{d}$} &
\multicolumn{2}{l}{$\bar{q} = \bar{s}$} \\ \hline
$J_{2j}^P$ & Mass & Decay & Mass & Decay \\
&$({\rm GeV}/c^2)$ & mode(s) & $({\rm GeV}/c^2)$ & mode(s) \\
&&&& \\
$0_1^+ $ & 5.68 & $(\overline{B} \pi )_{\ell = 0}$ & 5.78 &
$(\overline{B}K)_{\ell = 0}, \overline{B}_s^* \gamma $ \\
&&&& \\
$1_1^+$ & 5.68 & $(\overline{B}^* \pi )_{\ell = 0}$ & 5.78 &
$\overline{B}_s \gamma , ~ \overline{B}_s^* \gamma $ \\
&&&& \\
$1_3^+ $ & 5.78 & $ (\overline{B}^* \pi )_{\ell = 2}$ & 5.88 &
$(\overline{B}^* K)_{\ell = 2} $ \\
&&&& \\
$2_3^+$ & 5.79 & $(\overline{B} \pi )_{\ell = 2}, (\overline B^* \pi)_{\ell =
2}$ & 5.89 & $(\overline{B} K)_{\ell = 2},~(\overline{B}^* K)_{\ell = 2}$ \\
\hline
\end{tabular}
\end{center}

\medskip
\noindent
The $\ell=0$ decays [except for $b \bar{s} (0_1^+ ) \to \overline{B}K$,
which has very little energy release] should correspond to very broad
resonances, while the $\ell = 2$ decay widths should be tens of MeV or
less (as in the $D^*(2420)$ and $D^*(2460)$ cases).
Detailed estimates have been made in Ref.~\cite{Bres}.

\bigskip
\noindent
C.  {\bf The 2S states}
\bigskip

In order to make use of methods for tagging $D_s^+ = c \bar{s}$ or
$\overline{B}_s^0 = b \bar{s}$ using an associated kaon, one must study $K^-
D_s^+$ or $K^- \overline{B}_s^0$ combinations above threshold:  $2.46$ or $5.87
{}~ {\rm GeV}/c^2$, respectively. The $2_{3}^+ ~ c \bar{u}$ state, $D^*(2460)$,
should be just barely able to decay to $K^- D_s^+$. The $K^- \overline{B}_s^0$
threshold is above any of the nonstrange resonances in Table I.
If a resonance is to be responsible for $K^- \overline{B}_s^0$ or $K^-
\overline{B}_s^{*0}$ correlations, the lowest candidate will be a $2S$ state.

The spin\--weighted averages of $2S ~c \bar{c}$ and $b \bar{b}$ states probably
lie about $0.6~ {\rm GeV}/c^2$ above the corresponding $1S$ states. The spacing
between $1S$ and $2S$ states of one light quark and one heavy quark is probably
slightly greater than this \cite{AM,KQR}. In Ref.~\cite{Bres} the $2S - 1S$
spacings are found in a QCD-motivated potential of the Buchm\"uller-Tye
\cite{BT} type to be about (740, 720, 680, 660) MeV$/c^2$ for $(D,~B,~
D_s,~B_s)$ states. At any rate, the decay modes $\overline{B}_s^0 K^-$ and
$\overline{B}_s^{*0} K^-$ appear to be allowed for the $J^P = 1^-~2S~b\bar{u}$
state.

Making use of the estimates of Ref.~\cite{Bres} for nonstrange states but just
adding 100 MeV/$c^2$ for strange states, we expect the $2S~ c \bar{q}$ and $b
\bar{q}$ levels to have the approximate masses shown in Table II. If the
strange states really have smaller $2S - 1S$ spacings than the nonstrange ones,
as predicted in Ref.~\cite{Bres}, one should subtract about 60 MeV/$c^2$ from
the estimates in the second column of Table II.

\medskip

\begin{center}
\noindent
Table II.  Estimated masses and sample decay modes
of $2S~c \bar{q}$ and $b \bar{q}$ levels.

\medskip
\begin{tabular}{c | l l | l l |}
\multicolumn{3}{c }{$q = \bar{u}$ or $\bar{d}$} &
\multicolumn{2}{c |}{$q = \bar{s}$} \\

&&&& \\
& Mass & Decay & Mass & Decay  \\
$J^P$ & $({\rm GeV}/c^2)$ & mode(s) & $({\rm GeV}/c^2)$ & mode(s) \\ \hline
&&&& \\
$c \bar{q}~(0^-)$ & 2.68 & $D^* \pi,~D_s^* \bar K$ &  2.78 & $D^*K$ \\ \hline
&&&& \\
$c \bar{q}~(1^-)$ & 2.82 & $D^{(*)}\pi,~D_s^{(*)} \bar K$  &
2.92 & $D^{(*)}K$ \\ \hline
&&&& \\
$b \bar{q}~(0^-)$ & 6.00 & $\bar B^* \pi,~\bar B_s^* \bar K$ &
6.10 & $\bar B^* K$ \\ \hline
&&&& \\
$b \bar{q}~(1^-)$ & 6.05 & $\bar B^{(*)} \pi,~\bar B_s^{(*)} \bar K$ &
6.15 & $\bar B^{(*)}K$ \\ \hline
\end{tabular}
\end{center}

\bigskip

\noindent
Here we have assumed the same hyperfine splittings as in the $1S$ cases. The
hyperfine splitting in a nonrelativistic model should be proportional to $|
\Psi (0)|^2$, where $\Psi ({\bf r})$ is the Schr\"{o}dinger wave function. For
a system of reduced mass $\mu$ bound in a linearly rising potential $V(r) = ar
, ~ | \Psi (0) |^2 = ( \mu /4 \pi)$ $\langle dV/dr \rangle $  $= (\mu a /4 \pi
)$ independently of principal quantum numbers. There is some reason to suspect
that this is an appropriate limit for a light quark bound to a heavy one.

The results of Tables I and II suggest that $\bar B_s^{(*)} \bar K$
correlations may be similar to $D_s^{(*)} \bar K$ correlations, which should be
easier to study. One possible exception is that the $D(2460)$ should be just
barely to decay to $D_s^+ K^-$, while the corresponding $J^P = 2^+$ resonance
in the $B$ system is expected to be too light to decay to $\bar B_s K^-$.
\bigskip

\noindent
D.  {\bf Angular distributions and kinematics}
\bigskip

1.  {\it Effect of loss of photon in $B^* \to B \gamma$}.
The $D^*$ can decay to $D \pi$ or $D \gamma$, but the $B^*$ is only able to
decay to $B \gamma$. The energy of this photon is so low (about $46$ MeV) that
its detection is unlikely in most experiments. (See, however, Ref.~\cite{L3}.)
Even if the photon is missed in the decay $B^{**} \to B^* \pi \to B \gamma
\pi$, the effective mass of the $B \pi$ system is shifted down from the true
$B^{**}$ mass, but not broadened appreciably.

To see this, let $p_\pi , ~ p_B$ and $p_\gamma$ be the momenta of the pion,
$B$, and photon in the $B^*$ rest frame (Fig.~8). Let $\theta_\gamma$ be the
angle between the photon and the pion in this frame. We have $| \vec{p}_\gamma
| = E_\gamma = 46$ MeV $= | \vec{p}_B |$, while for $M(B^{**}) = 5.79$ GeV (the
value we predict for the $2_{3}^+$ state), one has $p_\pi = 464$ MeV. A bit
of arithmetic leads to
\begin{equation}
M_{B \pi} \simeq M_{B^{**}} - E_\gamma +
\frac{E_\gamma p_\pi}{M_{B^{**}}} \cos \theta_\gamma ~~,
\end{equation}
or, for $M(B^{**}) = 5.79$ GeV, $M(B \pi) \simeq M(B^{**}) - [46
- 3.8 (\cos \theta_\gamma)]~{\rm MeV}/c^2$. The predicted
mass differences between the $B \pi$ system and the $B$ are then:
\begin{equation}
M (B \pi ) - M(B) =
\left \{
\begin{array}{l r}
(448 + 4 \cos \theta_\gamma ) ~~{\rm MeV}/c^2 (1_{3}^+) &
{}~~~~~~~~~~ \\
(461 + 4 \cos \theta_\gamma ) ~~ {\rm MeV}/c^2 (2_{3}^+) &
{}~~~~~~~~~~ \\
\end{array}
\right .
\end{equation}
where in both cases a photon from $B^* \to B \gamma$ has been missed. Its loss
causes negligible broadening of the resonances.  The resonance masses are about
20 MeV/$c^2$ lower in the estimates of Ref.~\cite{Bres}. The decay of the
$2_{3}^+$ state to $B \pi$ leads to a peak with
\begin{equation}
M (B \pi) - M(B) \simeq 500 ~~ {\rm MeV}/c^2 ~~~~.
\end{equation}
The relative strengths of the peaks in $1_3^+$ and $2_3^+$ decay are
3:2 as shown in Refs.~\cite{cmts} and \cite{IW}.

\bigskip

2.  {\it Dalitz plot analysis of $D^{**} \to D^* \pi \to D \pi \pi$}.  Let us
define kinematic variables for the decays $D^{**} \to D^* \pi_1$, $D^* \to D
\pi_2$, as shown in Fig.~9. We recall some results already quoted in
Ref.~\cite{cmts} for the distribution in $\theta$ (equivalent to a Dalitz plot
variable). When a spin\--2 $D^{**}$ decays to $D^* \pi$, it does so via a
$D$\--wave, and the decay probability  $W (\theta )$, normalized in such a way
that
\begin{equation}
{1 \over 2} \int_{-1}^1 d (\cos \theta ) W ( \theta ) = 1 ~~~,
\end{equation}
is $W (\theta ) = (3/2) \sin^2 \theta$.
When a spin\--1 $D^{**}$ decays to $D^* \pi$, it can do so either by an
$S$\--wave (as expected for the $1_{1}^+$ state) or a $D$\--wave (as expected
for the $1_{3}^+$ state). The corresponding distributions are
\begin{equation}
W (\theta ) = \left \{
\begin{array}{c l r}
1 & (S ~~{\rm wave}) ~~, & ~~~~~~ \\
(1 + 3 \cos^2 \theta )/2 & (D ~~ {\rm wave}) ~~. & ~~~~~~ \\
\end{array}
\right .
\end{equation}
It appears that the decay $D (2420) \to D^* \pi$ is compatible with the
distribution for $D$ wave \cite{CLD,E687}. This supports the
identification of the $D(2420)$ as the $1_{3}^+$ state. The $D(2460)$ indeed
appears to have $J^P = 2^+$ \cite{ARGD,CLD}.

When and if another resonance decaying to $D^* \pi$ is discovered, we predict
that the distribution will be isotropic in $\theta$ as expected for the
$1_{1}^+$ state.
\bigskip

3.  {\it Dalitz plot analysis of $B^{**} \to B^* \pi \to B \gamma \pi$}. The
Dalitz plot distribution associated with the configuration noted in Fig.~8 can
be measured if one can detect the photon \cite{L3}.  Normalizing distributions
$W(\theta_\gamma)$ as above, we find for a spin-2 $B^{**}$ decaying to $B^*
\pi$, with subsequent decay of the $B^*$ to $\gamma B$, that $W(\theta_\gamma)
= 3(1 + \cos^2 \theta_\gamma)/4$.  This function is peaked at $\theta_\gamma =
0$ and $\pi$. The corresponding distributions for a spin-1 $B^{**}$ decaying to
$B^* \pi$ in a state of angular momentum $\ell$ are $W(\theta_\gamma) = 1$ for
$\ell = 0$ and $W(\theta_\gamma) = (2 + 3 \sin^2 \theta_\gamma)/4$ for $\ell =
2$.  This last function is peaked at $\theta_\gamma = \pi/2$.
\bigskip

4.  {\it Distributions for polarized $D^{**}$ and $B^{**}$}.
The Dalitz plot distributions corresponding to Fig.~9 cannot be measured for
$B^{**}$ decays since the decay $B^* \to B \pi$ is kinematically forbidden.
However, if $D^{**}$ or $B^{**}$ resonances are produced with any polarization,
their decays to $D^{(*)} \pi$ or $B^{(*)} \pi$ may produce pions with a
non\--isotropic distribution with regard to the polarization axis. This point
has recently been emphasized in Ref.~\cite{FP}.

Let us imagine that a spin\--J resonance $R$ (standing for $D^{**}$ or
$B^{**}$) is produced along some axis $\hat{n}$. By parity invariance one
expects the same probability for helicity $\lambda$ and $- \lambda$ with
respect to $\hat{n}$, but, aside from this, populations associated with
different helicities can differ. This, in turn, can lead to non\--trivial
distributions in the angle $\theta_1$ between the momentum of the pion $\pi_1$
to which the resonance $R$ decays and the direction $\hat{n}$. Labelling these
relative decay probabilties by $W_{| \lambda |} (\theta_1)$, where
\begin{equation}
{1 \over 2} \int_{-1}^1 d ( \cos \theta_1 ) W_{| \lambda | } ( \theta_1)
= 1 ~~~,
\end{equation}
\begin{equation}
W_0 (\theta_1 ) + 2 \sum_{| \lambda | > 0}^J W_{| \lambda | } ( \theta_1)
= 2J+1 ~~~,
\end{equation}
we have (for $P \equiv D$ or $B$, $V \equiv D^*$ or $B^*$):

\bigskip
\noindent
\underline{$R (2^+) \to P \pi$}
\begin{equation}
W_0 (\theta_1) = (5/4) (3 \cos^2 \theta_1 - 1 )^2
\end{equation}
\begin{equation}
W_1 (\theta_1) = (15/2) \sin^2 \theta_1 \cos^2 \theta_1
\end{equation}
\begin{equation}
W_2 (\theta_1 ) = (15/8) \sin^4 \theta_1
\end{equation}

\bigskip
\noindent
\underline{$R(1^+) \to (V \pi)_{\ell = 0}$}
\begin{equation}
W_0 (\theta_1 ) = W_1 (\theta_1 ) = 1
\end{equation}

\bigskip
\noindent
\underline{$R (1^+) \to (V \pi)_{\ell = 2}$}
\begin{equation}
W_0 (\theta_1 ) = (3/4) (1 + 3 \cos^2 \theta_1)
\end{equation}
\begin{equation}
W_1 (\theta_1 ) = (3/4) (1 + [3/2] \sin^2 \theta_1 )
\end{equation}

\bigskip
\noindent
\underline{$R (2^+) \to V \pi$}

\begin{equation}
W_0 (\theta_1) = (15/2) \sin^2 \theta_1 \cos^2 \theta_1
\end{equation}
\begin{equation}
W_1 (\theta_1) = (5/4) (1 - 3 \cos^2 \theta_1 + 4 \cos^4 \theta_1 )
\end{equation}
\begin{equation}
W_2 (\theta_1) = (5/4) (1 - \cos^4 \theta_1 )
\end{equation}
Of course, for $R(0^+) \to P\pi$ there is no $\theta_1$ dependence.

The above distributions are relevant to any attempt to select pion\--$D$ or
pion\--$B$ correlations by means of angular rather than effective\--mass cuts.
If different values of $| \lambda |$ are populated differently, such angular
cuts can either enhance or degrade a signal which was due originally to a
specific resonance or band of resonances.

\bigskip
\centerline{\bf VI.  CONCLUSIONS}
\bigskip

We have discussed the possibility of identifying neutral $B$ mesons using
hadrons produced nearby in phase space.  The simplest example is the expected
correlation between a $B^0$ and a $\pi^+$, which we expect to be stronger (with
relative probability $P_1$) than that between a $B^0$ and a $\pi^-$ (with
relative probability $P_2 < P_1$).  The correlation is expected to be most
pronounced for low effective masses or small rapidity differences.  It can
exist as a result of resonances in the $B \pi$ system, but can also be due
simply to the fragmentation of a $\bar b$ quark. All statements are of course
valid also for the charge-conjugate systems.

A number of issues have been treated in this article, which serves as a sequel
to Ref.~\cite{GNR}.

(1) We have noted some simple time-dependences in decays which are ``tagged''
by means of an associated hadron.  In general a dilution of the observed
asymmetry with a very simple form $(P_1 - P_2)/ (P_1 + P_2)$ occurs.

(2) Although we assume no coherence between $B^0$ and $\overline{B}^0$ in the
initial state, we have shown how to test for this coherence experimentally.

(3) We have stressed that explicit $B \pi$ resonances are not required for
``tagging,'' although the presence of such resonances may help to reduce
combinatorial backgrounds.

(4) We have mentioned the use of correlations with hadrons other than pions.
Quark diagrams describing fragmentation are particular helpful in visualizing
which correlations are likely to prove fruitful.

(5) We have stressed the need for detailed studies of the corresponding
correlations involving $D$ mesons, aside from the prominent production
of very soft pions in the decays $D^* \to D \pi$ which have no counterpart in
the $B$ system.

(6) We have treated several issues regarding resonances, discussing some
properties of the positive-parity charmed mesons and their extrapolation to
$B$ mesons, expected masses of $2S$ states, and angular distributions in
decays.

In the study of CP-violating decays of neutral $B$ mesons, the identification
of their initial flavor is a topic of keen interest.  The use of correlated
hadrons in this context is a promising possibility.  Whether it will be
realized in practice depends on a number of experimental questions, some of
which we have raised in the present work.
\bigskip

\centerline{\bf ACKNOWLEDGMENTS}

\bigskip

We are grateful to J.~Butler, E.~Eichten, H.~Frisch, G.~Gustafson, C.~Hill,
B.~Kayser, T.~LeCompte, H.~Lipkin, C.~Quigg, A.~I.~Sanda, V.~Sharma,
M.~Shochet, S.~Stone, and A.~Yagil for helpful discussions, and to D.~G.~Cassel
and H.~Tye for extending the hospitality of the Newman Laboratory of Nuclear
Science, Cornell University, during part of this investigation.  Part of this
study was performed at the Aspen Center for Physics.  This work was supported
in part by the United States - Israel Binational Science Foundation under
Research Grant Agreement 90-00483/2, by the Fund for Promotion of Research at
the Technion, and by the U.~S.~Department of Energy under Grant No. DE FG02
90ER-40560.

\newpage

\centerline{\bf FIGURE CAPTIONS}
\bigskip

\noindent
FIG. 1.  (a) Fragmentation of a $b$ quark into a $\bar B^0$ or $\bar B^{*0}$
with production of a $\pi^-$; (b) charge-conjugate process.
\bigskip

\noindent
FIG. 2.  (a) Fragmentation of a proton into a $b$-flavored baryon and (a) a
$B^+$ or (b) a $B^0$.
\bigskip

\noindent
FIG. 3.  Diagrams describing decays of a $B_s$ or $\overline{B}_s$ into $D_s^+
K^-$.
\bigskip

\noindent
FIG. 4.  Correlations of neutral nonstrange $B$ mesons with neutral $K^*$
resonances.  (a) $\bar B^0$ or $\bar B^{*0}$ with $K^{*0}$; (b)
charge-conjugate process.
\bigskip

\noindent
FIG. 5.  Correlations of strange $B$ mesons with charged kaons.
(a) $\bar B_s^0$ or $\bar B_s^{*0}$ with $K^-$; (b) charge-conjugate
process.
\bigskip

\noindent
FIG. 6.  Correlations of neutral nonstrange $B$ mesons with protons or
antiprotons.
\bigskip

\noindent
FIG. 7.  Correlations of charged kaons and charmed-strange mesons.
\bigskip

\noindent
FIG. 8.  Momenta of particles in the decay $B^{**} \to B^* \pi$, $B^* \to B
\gamma$, as expressed in the $B^*$ rest frame.
\bigskip

\noindent
FIG. 9.  Momenta of particles in the decay $D^{**} \to D^* \pi_1$, $D^* \to D
\pi_2$, as expressed in the $D^*$ rest frame.
\end{document}